\documentstyle[12pt,epsf]{article}
\newlength{\abstractwidth}
\renewcommand{\thefootnote}{\fnsymbol{footnote}}
\renewcommand{\thanks}[1]{\footnote{#1}} 
\newcommand{\starttext}{
\setcounter{footnote}{0}
\renewcommand{\thefootnote}{\arabic{footnote}}}
\renewcommand{\theequation}{\thesection.\arabic{equation}}
\newcommand{\be}{\begin{equation}}
\newcommand{\bea}{\begin{eqnarray}}
\newcommand{\eea}{\end{eqnarray}}
\newcommand{\beq}{\begin{equation}}
\newcommand{\ee}{\end{equation}}
\newcommand{\eeq}{\end{equation}}

\def\ba{\begin{eqnarray}}
\def\ea{\end{eqnarray}}

\def\12{{1 \over 2}}
\def\32{{3 \over 2}}
\def\72{{7 \over 2}}
\def\92{{9 \over 2}}

\def\nc{non--commutative}
\def\ny{non--commutativity}

\def\q{&=&}

\def\cs{Chern Simons Theory}

\def\qh{quasihole}

\def\qhs{Quantum Hall system}

\def\nccs{non--commutative Chern Simons Theory}
\def\pdag{\psi^{\dag}}

\begin{document}
\renewcommand{\theequation}{\thesection.\arabic{equation}}
\begin{titlepage}
\bigskip
\rightline{SU-ITP 01/35}
\rightline{hep-th/0107200}

\bigskip\bigskip\bigskip\bigskip

\centerline{\Large \bf {Realizing the Quantum Hall System
in String Theory }}

\bigskip\bigskip
\bigskip\bigskip

 \centerline{\it
 S. Hellerman and Leonard Susskind }
\medskip
\centerline{Department of Physics} \centerline{Stanford
University} \centerline{Stanford, CA 94305-4060}
\bigskip\bigskip
\begin{abstract}
In a recent paper Bernevig,  Brodie,  Susskind and Toumbas
constructed a brane realization of the Quantum Hall fluid.
Since then it has been realized that the Quantum Hall system
is very closely related to non--commutative Chern Simons theory
and this suggests alternative brane constructions which we believe
are more reliable and clear.
In this paper a brane construction is given for the
non--commutative Chern Simons Matrix formulation of the Quantum
Hall system as described by in recent papers by Susskind,
Polychronakos and by Hellerman and Van Raamsdonk. The system is a
generalized version of Berkooz's ``Rigid
Light Cone Membrane" which occurs as an excition of the DLCQ
description of the M5--brane in a background 3--form field.
The original construction of Berkooz corresponds to the fully
filled $\nu =1$ state of the QH system. To change the filling
fraction to $\nu = 1/(k+1)$ a system of $k$ background D8-branes
is required. Quasi--hole excitations can be generated by passing a
D6-brane though the Rigid Membrane.

\medskip
\noindent
\end{abstract}

\end{titlepage}
\starttext \baselineskip=18pt \setcounter{footnote}{0}

\setcounter{equation}{0}
\section{Chern Simons Matrix Theory and the Quantum Hall System }

According to \cite{susskind,polychronakos,hellerman, sakitaone, sakitatwo},
the quantum hall system at filling fraction $\nu = 1/(k+1)$ can be
described by Abelian \nc \ \cs \ at level $k$. \footnote{This Lagrangian
appears not to make sense for $k=0$ but we remind the reader that it
is a formal expression in which the definition of the level is
regulator-dependent.  In what follows we will use only the explicitly
matrix-regulated version.}
\be
S={k\over 4\pi} \int d^3y  \epsilon^{\mu \nu \lambda}
\left[
A_{\mu} \star \partial_{\nu} A_{\lambda}
+{2\over 3} A_{\mu} \star   A_{\nu} \star   A_{\lambda}
\right]
\ee
where the star--product is the usual Moyal product with \ny \
parameter $\theta$.  The parameters of the \qhs \ are the magnetic
field $B$ and the filling fraction $\nu$ given by
\bea
\nu \q 1/(k+1)  \cr
B \q (k+1)/\theta
\eea
Note that the connection between level and filling fraction is
slightly different than given in \cite{susskind}, ie $\nu = 1/k$.
The shift of $k$ by one is a quantum effect found by
Polychronakos.

 Alternatively, the theory
may be described by a matrix model involving classical Hermitian
matrix variables $X^i,A_0$. The index $i$ runs over the spatial
directions $i=1,2$ and $A_0$ is a matrix valued connection which
implements gauge invariance under unitary transformations in the
matrix space. The Lagrangian for the matrix theory is
\be
L=B Tr
\left\{
\epsilon_{ij} (\dot{X}^i +[A_0,X^i])X^j + 2\theta A_0
\right\}
\ee

The equation of motion for $A_0$  (Gauss law constraint) is
\be
[X^1,X^2] = i \theta
\ee

This equation can only be solved if the matrices are infinite
dimensional. This corresponds to an infinite number of electrons
on an infinite plane. There are many reasons to want to regulate
the system by taking the number of electrons to be finite, forming a finite
droplet of
Quantum Hall fluid with a boundary.
Polychronakos has given an elegant modification of the system
which accomplishes this \cite{polychronakos}. Following
Polychronakos we introduce a set of bosonic degrees of freedom
$\psi_n$ where the index $n$ runs over $(n= 1,2,...,N)$. The
matrices $X^i, A_0$ are now $N \times N$ Hermitian matrices. An
additional
term in the action is introduced
\be
L_{\psi} = \psi^{\dag}(i \dot{\psi} -A_0 \psi)
\ee

The Gauss law constraint becomes
\be
[X^1,X^2]=i\theta
\left(
I- {1 \over {k+1}}\psi \psi^{\dag}
\right)
\ee
where $I$ is the unit matrix and $\psi \psi^{\dag}$ represents the
matrix with components   $\psi_m \psi^{\dag}_n$. This equation no
longer requires infinite dimensional matrices. The constraint is
best understood in the following way. Take the trace to get
\be
\sum_m \psi_m \psi^{\dag}_m =N (k+1)
\ee
This expression is intended to be read as quantum-ordered and
tells us that there must be exactly $Nk$ quanta of the $\psi$
field present. These $Nk$ quanta reside at the boundary of the
droplet and provide the needed boundary degrees of freedom that
are implicit in a Chern Simons theory. The traceless part of the equation
is the $SU(N)$ generator and tells us that the state has to be
invariant under the operations
\be
X  \to u^{\dag}X u
\ee
This tells us that in forming states from the Fock space of the
oscillators $\psi, X$ we must contract all indices  to form
$SU(N)$ singlets. The states of this
system have been analyzed \cite{hellerman} and shown to be
in one to one correspondence with the states of the Laughlin
theory.

\setcounter{equation}{0}
\section{The Level Shift}

In the original paper on the QH system and
\nccs \ \cite{susskind} the connection
between level and filling factor was given as
\be
\nu = {1 \over k}.
\ee
Subsequently Polychronakos discovered a quantum correction
modifies this to
\be
\nu = {1 \over k+1}.
\ee
In this section we will give a simple derivation of this shift
based on the wave functions given in
\cite{hellerman}. The argument is due to Jeong-Hyuck Park and Dongsu Bak.
\cite{jeong}.

The first step is to find an operator in the matrix theory which
represents the area of the Quantum Hall droplet. For a uniform droplet
of
electrons it is easily seen that the right expression is
\be
Area={2 \pi \over N } \sum_n (X_n )^2.
\ee
The matrix analogue of this is
\be
Area= {2 \pi \over N } Tr (X)^2
\ee

Following \cite{hellerman} we define the $N \times N $ matrix of
harmonic oscillator operators
\be
A_{mn}\equiv \sqrt{{B\over 2}}(X^1+iX^2)_{mn}
\ee
The ground state found of the droplet is given by the state
\cite{hellerman}
\be
|k\rangle =
\left\{
\epsilon^{i_1--i_N}(\pdag)_{i_1}
(\pdag A^{\dag})_{i_2}
...(\pdag {A^{\dag}}^{N-1})_{i_N}
\right\}^k
|0\rangle
\ee

Now observe that $Tr (X)^2$ is given by
\be
Tr (X)^2 = {2 \over B}(Tr A^{\dag}A + \12 N^2  )
\ee
The expression
$Tr A^{\dag}A$ merely counts the total number of
$A^{\dag}$ that appear in (2.6).  The term $12 N^2$
is the zero-point fluctuation of the $N^2$ oscillators. This
zero-point oscillation is the cause of the level shift.

It is easily seen that for large $N$ the entire expression
becomes
\be
Tr(X)^2= {1 \over B} (k+1)N^2 
\ee
and from (2.4)
\be
Area={2 \pi \over B}(k+1)N
\ee
which corresponds to a filling fraction
$\nu = 1/(k+1)$.

The shift in the connection between level and filling factor
would seem to undo the relation between filling factor and
statistics found in \cite{susskind}. However there is a
compensating shift of statistics in matrix models that was also
overlooked in \cite{susskind}. The gauge invariant variables in an
$SU(N)$ invariant matrix model are the eigenvalues of the
matrices. The measure on the space of the eigenvalues involves a
so called Vandermonde determinant which may be absorbed into the
wave invariant functions. The result is to interchange Fermi
and Bose statistics
\footnote{This was explained to us by A. Polychronakos}.

In particular note that filling factor 1 is described by the
simplest possible theory with $k=0$!

\setcounter{equation}{0}
\section{Rigid Open Membranes}

The existence of a 5+1 dimensional  quantum field theory called
the (0,2) theory is essential for the consistency of string
theory. The theory may be thought of as the low energy description
of an M5-brane in 11 dimensional M-theory. The only concrete
construction of the (0,2) theory was given in \cite{aharony}
and consists of a DLCQ description obtained by considering
Matrix Theory \cite{BFSS} in the background of a
longitudinal 5-brane. In this description the elementary momentum
carriers are D0-branes.

The theory has also been studied in the background of a 3-form
field strength $H_{+ij}$  by Berkooz \cite{berkooz}  who finds that the
momentum
carriers blow up into ``rigid open membranes" with boundaries
on the 5-brane. The
process is similar to that by which strings in a background $B_{\mu \nu}$
field expand and form a dipole of size equal to their momentum
\cite{bigatti}. The effect is also a version of the Myers effect
\cite{myers}. In this paper we will see that Berkooz's Rigid Open
Membranes are ideal for modeling the Quantum Hall system \cite{bernevig}.

The system studied in \cite{berkooz} consists of a single M5-brane
wrapped on the compact light like direction $x^-$. The other
directions of the world volume are $x^+ ,X^1,X^2,X^3,X^4$. The
system may also be thought of as a D4-brane in 2a string theory.
The background $H$ field has components
\be
H_{+12}=H_{+34 } =H \neq 0
\ee

We will consider a DLCQ excitation carrying $N$ units of $P_-$.
That is
\be
P_- = N/R
\ee

In the appropriate decoupling limit
the corresponding matrix theory can be described in terms of
$N\times N$ matrices $X^i$ (i=1,2,3,4) and two complex
$N$-vectors $Q,\tilde{Q}$. The
$X^i$ transform as adjoints of $U(N)$ and the $Q^i$ as
fundamentals. The $X$ may be thought of as describing the strings
connecting the D0-branes and the $Q$ as describing strings
connecting the D0 and D4 branes. Following Berkooz we define
\bea
X \q X^1 +iX^2 \cr
\tilde{X} \q X^3+i X^4
\eea

The decoupling limit studied by Berkooz involves letting the 11
dimensional Planck mass $M_p$ and the field $H$  tend to infinity. In
this limit the Hamiltonian vanishes and the only equations of
motion which survives are  the vanishing of the D and F terms:
\bea
[X,X^{\dag}]+[\tilde{X},\tilde{X}^{\dag}]+
Q Q^{\dag}-(\tilde{Q})^{\dag}(\tilde{Q})
\q  \theta \cr
[X,\tilde{X}] + Q\tilde{Q}\q0
\eea
where $\theta$ is given by
\be
\theta = {H \over R M_p^6}.
\ee

If we specialize Berkooz's equations to the case of an open
membrane oriented in the $X^1,X^2$ plane then
\be
\tilde{X} =\tilde{Q} =0
\ee
and the equations take the form
\be
[X^1,X^2]+i Q Q^{\dag} =i\theta.
\ee
The important thing to notice is that this equation is the same as
eq(1.6) with the replacement
\be
Q \to \sqrt{\theta \over k} \psi.
\ee
Since there is no Lagrangian this system is identical to the $k=0$
version of Polychronakos' system. Thus Berkooz's Rigid open
membrane is a quantum Hall bubble with filling fraction $1$. The
term ``rigid" is being used by Berkooz in the same way as
``incompressible" is used in the Quantum Hall context.

\setcounter{equation}{0}
\section{Filling Fraction 1/n}

In the
limit $N \to \infty$ the D0-branes form an infinite 2-brane with a
distant boundary at infinity. Equivalently they form an infinite
Quantum Hall Droplet at filling fraction $\nu = 1$.
To change the filling fraction we need to introduce
something that will induce a Chern Simons term at level $k$ on the
membrane. Fortunately Brodie has told us how to do that \cite{brodie}.
Consider
adding a stack of D8-branes (in the type 2a description). The
D8-branes lie in the directions $X^1,X^2.....,X^8$ and are
displaced from the D4-brane along the $X^9$ axis. On one side of
the D8's the vacuum is the conventional flat vacuum of type 2a
string theory. On the other side there is a nonvanishing 10-form
field sourced by the D8's. In this region the vacuum is described
by massive 2a gravity.

Let us begin with the D4 and its
attached rigid open membrane on the trivial vacuum side. Now
transport $k$ of the D8's past the D4 system
\footnote{The system of
8-branes and 4-branes is BPS and therefore stable.
Adding the D0-branes leads to a non-BPS
configuration but will not destabilize the configuration
since it is a localized perturbation }.
Two important
effects occur, both of which are encoded in  Chern Simons terms
induced on branes in the massive 2a theory
\cite{green,brodie}. The induced term on a
$p$--brane  has the formal structure
\be
L_{p}=k A\wedge F^{p\over 2}.
\ee
In particular for a D0-brane the term is
\be
L_0=kA_0
\ee
which is just a chemical potential for string ends. It indicates
that the system formed from $N$ D0-branes must have $kN$
fundamental strings ending on it. This is also known as the Hanany
Witten effect \cite{hanany}. The other end of the string
can be on the 8-branes but it does not have to be. In fact if the
8-brane is moved well past the D4 system the stable configuration
will involve strings which end on the D4-brane. In other words we
will find that the number of $Q$ quanta will be $kN$ in exact
agreement with eq.(1.7).

Furthermore the rigid membrane will also have a Chern Simons
2-brane term induced which according to (4.1) will have the form
\be
L_2=kA\wedge F.
\ee
Actually this is only correct if there is no background $H$ field.
In the presence of the $H$ field ( B field in the 2a language) the
Chern Simons term must become \nc \ \cite{seiberg-witten}.
Evidently then, the effects induced on the open rigid membrane are
exactly what are needed to turn the system into that studied in
\cite{susskind,polychronakos,hellerman}. In other
words the system becomes the Quantum Hall System at filling
$\nu =1/(k+1)$.

\setcounter{equation}{0}
\section{Six-Branes and Quasiholes}

The D6-brane plays an interesting role in the brane/QH
correspondence. Recall that in Laughlin's theory the Quantum Hall
fluid will support quasiholes of fractional charge $\nu$. In
\cite{susskind} these \qh \ states were constructed by modifying
the Gauss law constraint (1.4) to allow an explicit source
\be
[X^1,X^2] = i\theta (1+ \nu P)
\ee
where $P$ is a projection operator of rank one in the matrix
space. If the projection operator projects onto a localized
coherent state in the oscillator representation of the matrix
space then the \qh \ is localized by the coordinates of the center
of the coherent state. From eq.(1.6) we see that
we can accomplish the same thing in the
regularized theory of \cite{polychronakos} by exciting a single $\psi$
quantum. In the brane representation this corresponds to adding an
additional string connecting the rigid open membrane to the
D4-brane.

Another way to view the \qh \ in the standard Quantum Hall
framework is to begin with a magnetic monopole on one side of the
plane containing the electrons. If we adiabatically pass the
monopole through the plane, say at the origin, the effect is to
push each electron to an orbit of one higher unit of angular
momentum. This leaves a hole at the origin which has a fractional
charge $\nu$.

The obvious candidate to replace the monopole in the 10
dimensional type 2a string theory is the D6-brane oriented in the
$X^3, ...,X^8$ direction. Suppose we pass the D6-brane through the
rigid open membrane piercing it at some location. If the
correspondence holds true it should create a \qh \ at that point.
In other words it should leave behind a string end on the
membrane. This is again an example of the Hanany--Witten effect
which requires just such a string to form when a D6-brane is
passed through a D2-brane.

After this work was completed we became aware of a similar brane
construction by Oren Bergman, John Brodie, and Yuji Okawa.  The setup that these
authors use is similar but not identical to the one reported here
and the conclusions generally agree \cite{bergman}.

\setcounter{equation}{0}
\section{Acknowledgements}
The authors would like to thank Mark Van Raamsdonk
for discussions and John Brodie and Oren Bergman
for communicating their results to us. L.S.
is especially grateful to Kimyeong
Lee, Piljin Yi, Dongsu Bak and Jeong-Hyuck Park for important
insights. L.S. would also like to thank Prof. C.W. Kim for
excellent hospitality at the Korea Institute for Advanced Study.

\end{document}